\documentclass[11pt,draftcls,onecolumn]{IEEEtran}
\usepackage[utf8]{inputenc}
\usepackage{amsmath,amssymb,amsthm}
\usepackage[noadjust]{cite}
\usepackage{multicol}

\usepackage{xspace}
\usepackage{fixltx2e}
\usepackage{graphicx,fancyhdr,epstopdf}
\usepackage[shortlabels]{enumitem}

\begin{document}
\title{Cell-Free Massive
MIMO with Nonorthogonal Pilots for Internet of Things}



\author{\IEEEauthorblockN{Shilpa Rao \\}
\IEEEauthorblockA{\textit{Center for Pervasive Communications} \textit{and
Computing, UC Irvine\\}}
\and
\IEEEauthorblockN{Alexei Ashikhmin \\}
\IEEEauthorblockA{\textit{Nokia Bell Labs,}
\textit{Murray Hill, NJ, USA\\}}
\and
\IEEEauthorblockN{Hong Yang \\}
\IEEEauthorblockA{\textit{Nokia Bell Labs,}
\textit{Murray Hill, NJ, USA}}
}

\maketitle
\begin{abstract}
We consider Internet of Things (IoT) organized on the principles
of cell-free massive MIMO. Since the number of things is very large, orthogonal
pilots cannot be assigned to all of them even if the things are stationary. This
results in an unavoidable pilot contamination problem, worsened by the fact that, for IoT,
since the things are operating at very low transmit power. To mitigate this
problem and achieve a high throughput, we use cell-free systems with
optimal linear minimum mean squared error (LMMSE) channel estimation, while traditionally simple suboptimal
estimators have been used in such systems. We further derive the analytical uplink and downlink signal-to-interference-plus-noise ratio (SINR)
expressions for this scenario, which depends only on large scale fading
coefficients. This allows us to design new power control algorithms that require
only infrequent transmit power adaptation. Simulation results show a 40\%
improvement in uplink and downlink throughputs and 95\% in energy efficiency over existing cell-free wireless systems and at least a three-fold uplink improvement
over known IoT systems based on small-cell systems.
\end{abstract}
\section{Introduction}
Internet of Things (IoT) represents an entirely new challenge to the wireless physical layer. In an IoT connectivity scenario, it is required to connect a large number, e.g., ten thousand or more,  of users (things) of which only a small fraction is active simultaneously. (In what follows we will use ``users" and ``things" interchangeably when we talk about IoT.) Challenges that are inherent to IoT include: scale (the number of users is significantly greater than in cellular wireless), low power requirements for things (things are anticipated either to use energy harvesting or infrequently replaced batteries), emphasis on the uplink, typically low and/or sporadic data outputs from each thing, and possibly ultra-low latency requirements. Wireless massive multiple-input multiple-output (MIMO) systems look to be one of the best candidates for resolving the above challenges.

A number of works exist on the use of massive MIMO for massive connectivity in IoT. In~\cite{liu2018massive} and~\cite{liu2018massive2}, the authors provide a framework for user activity detection and channel estimation with a cellular base station (BS) equipped with a large number of antennas, and characterize the achievable uplink rate. The user activity detection is based on assigning a unique pilot to each user, which serves as the user identifier. These pilots then are used
as columns of a sensing matrix. Active users synchronously send their pilots and a base station receives a linear combination of the those pilots. Next the base station runs a compressive sensing detection algorithm for the sensing matrix and identifies the pilots that occur  as terms in the linear combination. These pilots, in their turn, reveals the active users.

Massive MIMO for IoT connectivity in the context of cyber-physical systems (CPS) is considered in~\cite{lee2018massive}, and orthogonal pilot reuse to simultaneously support a large number of active industrial IoT devices is studied in~\cite{lee2020massive}. However, these works consider a centralized massive MIMO architecture where all the service antennas are collocated.
The cell-free architecture assumes that access points (APs) are distributed over
a wide area and that they are connected via a backhaul to a central processing
unit (CPU). The coherent processing across APs provided by this architecture
enables simple signal processing and power control.
Although the low backhaul requirement of centralized MIMO is advantageous, the
cell-free architecture is more suitable for IoT since it offers a greater
coverage area.

In~\cite{Hien_CF} and~\cite{nayebi2017precoding}, achievable
rates and power control algorithms for cell-free massive MIMO with a suboptimal channel estimation are
studied. This estimation is optimal only if orthogonal pilots are reused among the users. One of the reasons for choosing this suboptimal channel estimation was that it was a common understanding that the linear minimum mean square error (LMMSE) channel estimation allows estimation of the user signal-to-interference-plus-noise ratio (SINR) only as a function of instantaneous channel state information (CSI), i.e., small scale channel coefficients. At the same time it is very desirable to get SINR estimates that depend only on large scale channel fading coefficients, which do not depend on orthogonal frequency-division multiplexing (OFDM) tone index and change about 40 times slower then small scale fading coefficients.
Such expressions greatly reduce the complexity of power control and simplify the analysis of the system performance.
In this work we show that this common understanding  was a misconception, and derive uplink and downlink SINR expressions for the case of LMMSE channel estimation that are functions of only large scale fading coefficients, transmit power coefficients, and the used pilots. Our simulation results demonstrate that LMMSE channel estimation allows obtaining an additional $40\%$ performance gain in terms of data transmission rates compared with suboptimal channel estimation used in \cite{Hien_CF} and \cite{nayebi2017precoding}. We next propose efficient power control algorithms. In the uplink, power control is performed by considering two criteria -- the max-min SINR criterion and a target SINR criterion. In the latter, we try to ensure that each user attains a pre-defined SINR threshold. In the downlink, the power control optimization problem is formulated as a second-order cone program (SOCP) that can be solved efficiently. We also compare our systems with small-cell systems, which have been suggested as promising technology for future wireless systems~\cite{liu2014comparative,liu2014energy}. In a typical small-cell system, the APs are uniformly distributed in the coverage area and each user is served by a dedicated AP. Since the APs operate independently, the average data rate in a small-cell system is lower than in a cell-free system. We apply our power control algorithms to the system in~\cite{Hien_CF} under correlated and uncorrelated shadow fading channels and show a significant gain in uplink and downlink throughputs over results obtained in~\cite{Hien_CF}, and over conventional small-cell systems.

In section~\ref{sec:SysModel}, we describe the cell-free massive MIMO system model, channel estimation, and the uplink and downlink transmissions. In section~\ref{UpSINR}, the uplink SINR when the APs use matched filtering based on the LMMSE channel estimate is derived. We propose two uplink power control algorithms in section~\ref{UpPower}. In section~\ref{DownSINR}, the downlink SINR is derived and in section~\ref{DownPower}, the downlink power control is performed. Then, we summarize the results for small-cell systems in section~\ref{SmallCell}. Finally, the simulation results are presented in section~\ref{Sims}.

\emph{Notation}: Boldface lowercase variables denote vectors and boldface uppercase variables denote matrices. ${\bf X}^T$, ${\bf X}^H$ and ${\bf X}^*$ are the transpose, Hermitian transpose, and conjugate of ${\bf X}$, respectively. The $i$th element of vector ${\bf x}$ is represented by ${x}_i$ and the $(i,j)$th element of matrix ${\bf X}$ is denoted by $x_{ij}$. $\mathbb{E}(\cdot)$ is the expectation operator and ${\bf X}^{-1}$ denotes the inverse of ${\bf X}$. The matrix ${\bf I}_p$ denotes a $p \times p$ identity matrix. A circularly symmetric complex Gaussian random variable with mean $a$ and variance $b$ is denoted by ${\bf x} \sim \mathcal{CN}\left( a, b\right)$. 

\section{System Model}
\label{sec:SysModel}

\subsection{Channel Estimation}\label{subsec:ChannelEst}
We consider a wireless system with $N$ things, among which only $K$ are
active at any given moment, and $M$ APs that are connected via a backhaul to a CPU. We assume that a unique pilot is assigned to each user and active things are detected by each AP, e.g., by using the approach proposed in \cite{liu2018massive}.
Without loss of generality, and to simplify
notation, we assume that things $1,\ldots,K$ are active. We assume that OFDM is used and, consequently, we consider a flat-fading
channel model for each OFDM subcarrier. For a given subcarrier
 we model the {\em channel
coefficient} $g_{mk}$ between the $k$-th user and the $m$-th AP as
\begin{align*}
g_{mk} = \sqrt{\beta_{mk}} h_{mk},
\end{align*}
where $h_{mk} \sim \mathcal{CN}(0,1)$ is the {\em small-scale fading coefficient} and
$\beta_{mk}$ is the {\em large-scale fading coefficient} that includes path loss and
shadowing coefficient. We assume that $h_{mk}$ are i.i.d. ${\cal CN}(0,1)$ that they stay constant during
the coherence interval of duration  $\tau_c$ OFDM symbols. Note that for different OFDM tones we have different $h_{mk}$-s.
In contrast,  large-scale fading coefficient $\beta_{mk}$ do not depend on frequency, i.e., they are the same for all OFDM tones.
They also change typically about $40$ times slower than coefficients $h_{mk}$. For this reason the coefficients $\beta_{mk}$ can
be accurately estimated and therefore we treat them as known constants known to all APs. With these assumptions we obtain that
$g_{mk}$ are independent Gaussian random variables.

Let ${\bf g}_m = [ g_{m1}, \dots,g_{mK} ]^T$ be the {\em channel vector} and
${\bf G} \in \mathbb{C}^{M \times K}$ be the {\em channel matrix} with $[{\bf G}]_{m,k} = g_{mk}$. During
pilot transmission, the $K$ active users synchronously transmit {\em pilot sequences} $\boldsymbol{\psi}_j\in \mathbb{C}^{\tau}$ of length $\tau$. Let  $\boldsymbol{\Psi} =[
\boldsymbol{\psi}_1 \, \boldsymbol{
\psi}_2 \, \dots \, \boldsymbol{\psi}_K ] \in \mathbb{C}^{\tau \times K}$. Thus, the received signal at the APs, ${\bf Y} \in \mathbb{C}^{\tau \times M}$, is given by
\begin{equation*}
{\bf Y} = [{\bf y}_1\, {\bf y}_2 \, \dots \, {\bf y}_M \ ] = \sqrt{\tau \rho_p} \boldsymbol{\Psi} {\bf G}^T + {\bf W},
 \label{Eq2}
\end{equation*}
where $\rho_p$ is the {\em pilot transmission power} and ${\bf W} = \left[ {\bf w}_1, {\bf w}_2, \dots, {\bf w}_M \right]$ is the additive noise matrix with i.i.d. entries  $w_{ij} \sim \mathcal{CN}(0, 1)$.
Then
\begin{equation}\label{eq:y[m]}
{\bf y}_m=\sqrt{\tau \rho_p} \boldsymbol{\Psi}{\bf g}_{m}+{\bf w}_{m}.
\end{equation}
The $m$-th AP computes the LMMSE estimate $\hat{\bf g}_m$ of  ${\bf g}_m$ as follows
\begin{equation*}
\begin{aligned}
\hat{\bf g}_{m} = & \mathbb{E} [ {\bf g}_m {\bf y}^H_m ] ( \mathbb{E} [{\bf y}_m {\bf y}^H_m ] )^{-1} {\bf y}_m \\
= &  \sqrt{\tau \rho_p} {\bf B}_m \boldsymbol{\Psi}^H ( \tau \rho_p \boldsymbol{\Psi}  {\bf B}_m \boldsymbol{\Psi}^H + {\bf I}_{\tau}  )^{-1} {\bf y}_m, \,\,\, m=1,\dots,M,
\end{aligned}
\label{Eq3}
\end{equation*}
where ${\bf B}_m = {\rm diag} \{ \beta_{m1},\beta_{m2},\dots,\beta_{mK} \} $. It is convenient to  define ${\bf A}_m, m=1,\dots,M,$ and express  $\hat{{\bf g}}_m$ in terms of ${\bf A}_m$ as
\begin{equation}
\begin{aligned}
 {\bf A}_m= & \sqrt{\tau \rho_p} ( \tau \rho_p \boldsymbol{\Psi}  {\bf B}_m \boldsymbol{\Psi}^H + {\bf I}_{\tau}  )^{-1} \boldsymbol{\Psi}  {\bf B}_m \in {\mathbb C}^{\tau \times K}, \\
\hat{\bf g}_{m} = & [ \hat{g}_{m1}, \dots,\hat{g}_{mK} ]^T = {\bf A}^H_m  {\bf y}_m.
\end{aligned}
\label{Eq4}
\end{equation}
Using (\ref{eq:y[m]}) and~(\ref{Eq4}), we obtain that the covariance of $\hat{\bf g}_{m}$ is
\begin{align*}
\mathbb{E} [ \hat{\bf g}_{m} \hat{\bf g}_{m}^H ]
=&{\bf A}_m^H \mathbb{E} [
(\sqrt{\tau \rho_p} \boldsymbol{\Psi}{\bf g}_{m}^T+{\bf w}_{m})(\sqrt{\tau \rho_p} \boldsymbol{\Psi}{\bf g}_{m}^T+{\bf w}_{m})^H
]{\bf A}_m\\
=& {\bf A}_m^H ( \tau \rho_p \boldsymbol{\Psi}  {\bf B}_m \boldsymbol{\Psi}^H + {\bf I}_{\tau}  ) {\bf A}_m
= \sqrt{\tau \rho_p} {\bf B}_m  \boldsymbol{\Psi}^H {\bf A}_m,
\end{align*}
and the variance of the estimate $\hat{g}_{mk}$ is equal to
\begin{align*}
\gamma_{mk}  \triangleq \mathbb{E} [ |\hat{g}_{mk}|^2 ] = \sqrt{\tau \rho_p} \beta_{mk}  \boldsymbol{\psi}_k^H {\bf a}_{m,k}, \label{Eq5}
\end{align*}
where ${\bf a}_{m,k}$ is the $k$-th column of ${\bf A}_{m}$.
We denote the {\em channel estimation error} by
 $\tilde{g}_{mk}$ and note that $ \tilde{g}_{mk} \sim \mathcal{CN}(0, \beta_{mk} - \gamma_{mk} )$ and that $\tilde{g}_{mk}$ and
 $\hat{g}_{mk}$ are uncorrelated.

\subsection{Uplink and Downlink Data Transmissions}\label{sec:ULandDL}
During the uplink transmission, all $K$ active users simultaneously send their data that propagate to all APs. Each user uses a {\em uplink power coefficient} $\eta_k$ to weigh its transmitted symbol $s_k$, where $\mathbb{E}[ |s_k|^2 ] =1$. The received signal at the $m$-th AP, ${y}^u_{m}$, is given by
\begin{equation*}
\begin{aligned}
{y}^u_{m} = \sqrt{\rho_u} \sum_{j=1}^K \sqrt{\eta_j} g_{mj} s_j + w^u_m,
\end{aligned}
\label{Eq6}
\end{equation*}
where $\rho_u$ is the {\em maximum uplink transmit power} and $w^u_m$ is the additive noise. We assume that $w^u_m,m=1,\ldots,M$, are i.i.d. ${\cal CN}(0,1)$ random variables. The uplink power coefficients $\eta_k$ should satisfy the constraints $0 \leq \eta_k \leq 1, k=1,\dots,K$. The $m$-th AP sends the products ${y}^u_{m} \hat{g}^*_{mk},~\forall k$, to the CPU.
To estimate  the symbol from the $k$-th user, the CPU adds the products  received from all the APs to get the estimate
\begin{equation}
\begin{aligned}
\hat{s}_k = &  \sum_{m=1}^M \hat{g}^*_{mk} {y}^u_{m}
=  \sqrt{\rho_u} \sum_{j=1}^K \sum_{m=1}^M \sqrt{\eta_j}  \hat{g}^*_{mk}  g_{mj} s_j + \sum_{m=1}^M  \hat{g}^*_{mk} w^u_m.
\end{aligned}
\label{Eq7}
\end{equation}

In the downlink, the APs use the channel estimates to perform conjugate
beamforming and transmit data to the $K$ active users. Each AP uses {\em downlink power
  coefficients} $\eta_{mk}$ to weigh the symbol intended for the $k$-th
user, $s_k, k=1,\dots,K$. The transmitted signal from the $m$-th AP, ${x}^d_{m}$,
is given by
\begin{equation}
\begin{aligned}
{x}^d_{m} = \sqrt{\rho_d} \sum_{j=1}^K \sqrt{\eta_{mj}} \hat{g}^*_{mj} s_j,
\end{aligned}
\label{Eq6D}
\end{equation}
where $\rho_d$ is the {\em maximum downlink transmit power}. The transmitted signals
should satisfy the constraints $\mathbb{E}[|  x^d_m
|^2 ] \leq \rho_d,\,\, m=1,\dots,M$. Thus, $\sum_{j=1}^K \eta_{mj}
\gamma_{mj} \leq 1$. Let $w_k^d \sim \mathcal{CN}(0,1)$ be the additive noise at
the $k$-th user. Then, the received signal at the $k$-th user is
\begin{equation}
\begin{aligned}
y^d_k = & \sum_{m=1}^M g_{mk} x_m^d + w_k^d =  \sqrt{\rho_d} \sum_{m=1}^M
\sum_{j=1}^K \sqrt{\eta_{mj}} g_{mk}  \hat{g}^*_{mj} s_j  + w_k^d.
\end{aligned}
\label{Eq7D}
\end{equation}

\section{Uplink SINR and Energy Efficiency}
\label{UpSINR}
In the uplink, the CPU detects symbol $s_k$ as $\hat{s}_k$, which can be written as
\begin{align}
\hat{s}_k =  &   s_k \underbrace{ \sqrt{\rho_u \eta_k} \sum_{m=1}^M
\mathbb{E}[ \hat{g}^*_{mk}  g_{mk} ] }_{T_1} + \underbrace{
\sqrt{\rho_u \eta_k} s_k \sum_{m=1}^M ( \hat{g}^*_{mk}  g_{mk} -
\mathbb{E}[ \hat{g}^*_{mk}  g_{mk} ] )}_{T_2}
\nonumber \\
& \,\,  + \underbrace{ \sum_{i \neq k} \sqrt{\rho_u \eta_i} s_i \sum_{m=1}^M  \hat{g}^*_{mk} g_{mi} }_{T_3} + \underbrace{\sum_{m=1}^M  \hat{g}^*_{mk} w^u_m}_{{\rm T_4}}.\label{Eq8}
\end{align}

In (\ref{Eq8}), the square of $T_1$ is the signal power.  The variances of $T_2$, $T_3$, and $T_4$ are the powers of the beamforming uncertainty error and channel estimation error, the pilot contamination error and interference, and the additive noise respectively. Let $ g_{mk} =   \hat{g}_{mk} + \tilde{g}_{mk}$, where $\hat{g}_{mk}$ and $\tilde{g}_{mk}$ are defined in Section \ref{subsec:ChannelEst}.
We assume that the CPU has the channel statistics in its possession. In order to derive the uplink SINR, we use the following lemmas:

\emph{Lemma 1}: $\mathbb{E}[\hat{g}_{mk} \tilde{g}_{mk}] = 0$.

\emph{Proof}: This property follows from the orthogonality of the LMMSE estimate and the estimation error. \qed

\emph{Lemma 2}: For $m \neq n$, $\mathbb{E}[\hat{g}_{mk} \hat{g}_{nk}^*] = 0$.

\emph{Proof}: If $r\not =s$, then, according to the system model, $g_{rk}$ and $g_{sk}$ are uncorrelated for any $k$ and therefore
 $\mathbb{E}[{\bf g}_{r}{\bf g}_{s}^H]={\bf 0}$. Since we also have $\mathbb{E}[{\bf g}_{r}{\bf w}_{s}^H]={\bf 0}$
 and $\mathbb{E}[{\bf w}_{r}{\bf w}_{s}^H]={\bf 0}$, we obtain
 \begin{align*}
\mathbb{E}[\hat{g}_{mk} \hat{g}_{nk}^*]&=\mathbb{E}[{\bf a}_{m,k}^H {\bf y}_{m} {\bf y}_{n}^H{\bf a}_{n,k}]\\
&={\bf a}_{m,k}^H \mathbb{E}[(\sqrt{\tau\rho_p} {\bf \Psi}{\bf g}_{m}+{\bf w}_{m})
(\sqrt{\tau\rho_p} {\bf \Psi}{\bf g}_{n}+{\bf w}_{n})^H] {\bf a}_{n,k} = 0.
\end{align*}
 \qed

\emph{Lemma 3}: $\mathbb{E}[| \hat{g}_{mk} |^4] = 2 \gamma_{mk}^2$.

\emph{Proof}: Let $\hat{g}_{mk} = \sum_{l} a_l z_l$, where the $a_l$ are constants and $z_l$ are i.i.d random variables with $z_l \sim \mathcal{CN}(0,1)$. Then, $\sum_l a_l^2 = \gamma_{mk}$. Furthermore, we have that $\hat{g}_{mk} = b_1 + j b_2,\ j=\sqrt{-1}$, where
 $b_1$ and $b_2$ are uncorrelated $\mathcal{N}(0,\sum_l a_l^2/2)$ random variables. Hence, $| \hat{g}_{mk} |^4 = (b_1^2 + b_2^2)^2 = b_1^4 + b_2^4 + 2 b_1^2 b_2^2$. Then, $\mathbb{E}[| \hat{g}_{mk} |^4] = \frac{3}{4} \gamma^2_{mk} +\frac{3}{4} \gamma^2_{mk} +2 \frac{\gamma_{mk}}{2} \frac{\gamma_{mk}}{2} = 2 \gamma_{mk}^2$. \qed

From these lemmas it follows that $T_2,T_3$, and $T_4$ are mutually uncorrelated, and that they are also uncorrelated with signal $s_k$.
Thus the sum of their variances is the power of the ``effective noise", and the SINR of the $k$-th user is
\begin{equation*}
\begin{aligned}
{\rm SINR}_k = & \frac{ | T_1 |^2  }{{\rm Var}[T_2]+{\rm Var}[T_3]+{\rm Var}[T_4] }.
\end{aligned}
\label{Eq15preA}
\end{equation*}
This leads to the following result.

\emph{{\bf Theorem 1}}: An achievable uplink data transmission rate of  the $k$-th user in
cell-free Massive MIMO with LMMSE channel estimation and matched filtering receiver is
\begin{equation}\label{R_k^u,cf}
R_k^{u,cf} = \log_2 (1+{\rm SINR}_k),
\end{equation}
 where
{\small
\begin{align}
&{\rm SINR}_k  \nonumber \\
= & \frac{ \rho_u \eta_k ( \displaystyle \sum_{m=1}^M \gamma_{mk} )^2} {  \displaystyle \sum_{m=1}^M \gamma_{mk} ( 1+ \rho_u \eta_k  \beta_{mk} ) +  \rho_u \displaystyle  \sum_{i \neq k} \eta_i [ \displaystyle \sum_{m=1}^M \beta_{mi}   \lVert {\bf a}_{m,k} \rVert^2_2 +   \tau \rho_p ( | \displaystyle \sum_{m=1}^M \beta_{mi}  \boldsymbol{\psi}^H_i {\bf a}_{m,k} |^2+ \displaystyle \sum_{m=1}^M \sum_{j=1}^K \beta_{mi} \beta_{mj} | \boldsymbol{\psi}^H_j {\bf a}_{m,k} |^2 ) ] }\label{Eq15}
\end{align}
}

\emph{{\bf Proof}}:  The effective noise $T_2+T_3+T_4$ is not Gaussian. However, from the fact that $s_k$ is uncorrelated with the effective noise, using \cite{hassibi2003much}, we obtain the lower bound on the mutual information
$$
I(\hat{s}_k, s_k) \ge {\rm log}_2(1+ {| T_1 |^2 \over
{\rm Var}[T_2]+{\rm Var}[T_3]+{\rm Var}[T_4]}
  ).
$$
Derivations of the variances are presented in Appendix A. \qed

\emph{Remark: Orthogonal pilot scenario}.

It is instructive to consider the case of orthonormal pilots, i.e., $\boldsymbol{\Psi}^H \boldsymbol{\Psi} = {\bf I}_K$.
In this case  we have
\begin{equation*}
\begin{aligned}
{\bf A}_{m} \overset{{\rm (a)}}{=} &\sqrt{\tau \rho_p}  \boldsymbol{\Psi} {\bf B}_m  (  \tau \rho_p {\bf B}_m  + {\bf I}_K  )^{-1},~
{\bf a}_{m,k} = \frac{\sqrt{\tau \rho_p}  \beta_{mk} \boldsymbol{\psi}_k }{1+ \tau \rho_p \beta_{mk}},\mbox{ and }
\gamma_{mk}={\tau\rho_p \beta_{mk}^2\over 1+\tau\rho_p\beta_{mk}},
\end{aligned}
\label{EqRem1}
\end{equation*}
where in (a) we have used the identity $( {\bf I} + {\bf A} {\bf B} )^{-1} {\bf A} = {\bf A} ( {\bf I} +  {\bf B} {\bf A})^{-1} $. This leads to the following observations:
\begin{equation}
\begin{aligned}
&\mathbb{E}[\hat{{\bf g}}_m \hat{{\bf g}}_m^H ] =  \tau \rho_p {\bf B}_m^2 (  \tau \rho_p {\bf B}_m + {\bf I}_K )^{-1}, \\
&\lVert  {\bf a}_{m,k} \rVert_2^2 = \frac{\tau \rho_p \beta_{mk}^2 }{(1+ \tau \rho_p \beta_{mk})^2} = \frac{\gamma_{mk}}{1+ \tau \rho_p \beta_{mk}}, \\
& \boldsymbol{\psi}_k^H {\bf a}_{m,k} = \frac{\sqrt{\tau \rho_p} \beta_{mk}}{1+
\tau \rho_p \beta_{mk}}, \mbox{ and }{\rm if \,} i \neq k, \quad \boldsymbol{\psi}_i^H {\bf a}_{m,k} = 0.
\end{aligned}
\label{EqRem2}
\end{equation}
Hence, (\ref{Eq15}) becomes
\begin{equation}
\begin{aligned}
{\rm SINR}_k =  \frac{ \rho_u \eta_k ( \sum_{m=1}^M \gamma_{mk} )^2} {  \sum_{m=1}^M \gamma_{mk} + \rho_u \sum_{i=1}^K \eta_i \sum_{m=1}^M  \gamma_{mk} \beta_{mi}  },
\end{aligned}
\label{EqRem3}
\end{equation}
\normalsize
which corresponds to the SINR in (28) of \cite{Hien_CF}, where the pilots are orthonormal and the channel estimation is obtained as
$$
\hat{g}_{mk}={\sqrt{\tau\rho_p} \beta_{mk} \over \tau\rho_p \beta_{mk}+1} \psi_k^H {\bf y}_m.
$$
 Furthermore, when the APs are collocated, $\beta_{mk} \triangleq \beta_{k}$, $\gamma_{mk} \triangleq \gamma_k$, and (\ref{EqRem3}) simplifies to
\normalsize
the uplink SINR expressions obtained in~\cite{Hien_largeMU,yang_large}.

In IoT systems, unlike conventional wireless communications systems, not only high SINRs, equivalently spectral efficiency, is important. In a IoT system things will use energy harvesting and/or
infrequently replaced batteries. For this reason another important characteristic is the {\em energy efficiency} defined by
\begin{equation}\label{eq:ee}
E_u = {\sum_{k=1}^K R_k^u\over P_u \sum_{k=1}^K \eta_k},
\end{equation}
where $R_k^u$ is the uplink data rate of the $k$-th user and $P_u$ is the maximum transmission power for each thing in the uplink. For cell-free massive MIMO, the uplink data rate is given by~(\ref{R_k^u,cf}).

\section{Uplink power control}
\label{UpPower}
In order to provide good service to all users, it is important to conduct optimization for the uplink power coefficients, $\eta_k, k=1,2,\dots,K$. We assume that the CPU uses Theorem 1 to find optimal $\eta_k,~\forall k$, and to communicate them back to the APs, which forward these coefficients to the users.
Since the SINR expression in~(\ref{Eq15}) is a function of large-scale fading channel parameters and the pilot symbols, the power control is performed on a large-scale fading time scale. Moreover, since large scale fading coefficients do not depend on OFDM tone index, for each user $k$, it is enough to find only one $\eta_k$. This greatly reduces the amount of data that should be communicated from CPU to the users.

In what follows, we consider two criteria for power control.  The first one is the commonly used max-min criterion in which we seek to maximize the minimum of the uplink rates of all the users in order to guarantee a uniform service to all users. The second is the target SINR criterion where the aim is to ensure that each user attains a pre-defined SINR threshold. The latter optimization can be performed in a distributed manner.
\subsection{Max-min power control}
In the uplink, the max-min power control problem can be stated as follows:
\begin{equation*}
\begin{aligned}
& {\rm max}_{\{ \eta_k \}} \,\, {\rm min}_{k=1,\dots,K} \,\, {\rm SINR}_k \\
& \text{subject to}\,\, 0 \leq \eta_k \leq 1, \,\, k=1,\dots,K.
\end{aligned}
\label{EqUPC1}
\end{equation*}
This optimization problem can be reformulated as
\begin{equation}
\begin{aligned}
 &{\rm max}_{\{ \eta_k \}, t} \,\,  t  \\
 &\text{subject to}\,\,  t \leq {\rm SINR}_k, \,\, k=1,2,\dots,K, \\
 & \hspace{15mm} 0 \leq \eta_k \leq 1, \,\, k=1,2,\dots,K.
\end{aligned}
\label{EqUPC2}
\end{equation}
The optimization problem  (\ref{EqUPC2}) is quasi-linear and can be efficiently solved by the bisection search at each step of which we solve  a linear feasibility problem.
\subsection{Target SINR power control}
Max-min power control optimization is a centralized algorithm that guarantees a
uniform SINR to all the users. A downside is that if a user suffers from a bad
channel gain and experiences poor SINR, the achievable rates for all the other
users are compromised. Below, we propose a distributed algorithm based on the
algorithm developed in \cite{rasti2011distributed}, for optimal power control
with target SINRs varying among the users.\footnote{Strictly speaking, the
centralized method can also be used for the case with different user SINRs:
Given ${\rm SINR}_1,\dots, {\rm SINR}_K$, solve the linear equations for
$\eta_1,\dots,\eta_K$, and if they satisfy $0\le \eta_i\le 1$, ${\rm
SINR}_,\dots, {\rm SINR}_K$ are achievable.} This
approach is realistic for IoT where some things might transmit at a higher bit
rate than other things. The decentralized approach also allows for ``soft
removal", where some users are gradually removed in order to support users with
better channels. The method assumes that APs send the current SINR
values to the users, and that the users use this information to update their transmit powers. This
type of power control was used for centralized massive MIMO systems in
\cite{adhikary2017uplink}. Formally, we define the optimization problem as
\begin{equation}
\begin{aligned}
&{\rm min}  \quad \sum_{k=1}^K \eta_{k} \\
&\text{subject to}\quad {\rm SINR}_k \geq \delta_{k}, k=1,\dots,K\\
& \hspace{1.8cm} 0 \leq \eta_k \leq 1, k=1,\dots,K.
\end{aligned}
\label{EqUPC3}
\end{equation}
Here, $\delta_k$ is the target SINR for the $k$-th user. Assuming $\delta_k, k=1,
\cdots, K$ are attainable, Algorithm 1 (below) solves~(\ref{EqUPC3}).

\noindent{\bf Algorithm 1}
\begin{enumerate}
\item Let $\eta_k^{0} = 1, \,\,k=1,\dots,K$.
\item Assign $n=1$, repeat steps 3-4 until $| {\rm SINR}_k^n - \delta_k| < \epsilon, \forall k$, for an error tolerance level $\epsilon >0$.
\item The corresponding SINRs, ${\rm SINR}_k^{n-1}, \forall k, $ are computed (and sent to the users) and the new transmit power is estimated as
 $\eta_k^{n} =  \begin{cases}
\eta_k^{n-1} \frac{\delta_k}{{\rm SINR}_k^{n-1}}, \quad \text{if } \frac{\eta_k^{n-1}}{{\rm SINR}_k^{n-1}} \leq \frac{1}{\delta_k} \\
{\rm min} ( 1, \frac{\rho_u}{\delta_k} \frac{{\rm SINR}_k^{n-1}}{\eta_k^{n-1}}), \quad \text{otherwise}.
\end{cases}$
\item $n=n+1.$
\end{enumerate}

\emph{{\bf Theorem 2}}: The algorithm always converges and converges to the
optimal powers when~(\ref{EqUPC3}) is feasible.

\emph{{\bf Proof}}: See Appendix B.  \qed

\section{Downlink SINR expression}
\label{DownSINR}

Recall the assumption that each user knows large scale fading coefficients,
but does not have any estimates of small scale fading coefficients knowledge of
the channel. The symbol received by the $k$-th user is defined  in~(\ref{Eq7D})
and it can be written as
\begin{equation}
\begin{aligned}
y^d_k =  &  \underbrace{\sqrt{\rho_d} s_k \sum_{m=1}^M  \sqrt{\eta_{mk}}\,
\mathbb{E}[ \hat{g}^*_{mk}  g_{mk} ] }_{T_1} + \underbrace{
\sqrt{\rho_d} s_k \sum_{m=1}^M \sqrt{\eta_{mk}} ( \hat{g}^*_{mk}  g_{mk} -
\mathbb{E}[ \hat{g}^*_{mk}  g_{mk} ] )}_{T_2}
\\ & \,\,  + \underbrace{ \sum_{i \neq k} \sqrt{\rho_d} s_i \sum_{m=1}^M
\sqrt{\eta_{mi}}  \hat{g}^*_{mi} g_{mk} }_{T_3} +
\underbrace{w^d_k}_{\rm T_4}.
\end{aligned}
\label{Eq8D}
\end{equation}
Using an approach that is similar to the one we used for the uplink case, we get
a closed-form expression for the downlink SINR.

\emph{{\bf Theorem 3}}: An achievable downlink rate for the $k$-th user in
cell-free massive MIMO with LMMSE channel estimation and conjugate beamforming
is given by~(\ref{Eq15D}) shown below
\begin{equation}\label{Eq15D}
{\rm SINR}^d_k = N_k/D_k,
\end{equation}
where
\begin{align*}
N_k  = &\rho_d  ( \sum_{m=1}^M  \eta_{mk}^{1/2}
\gamma_{mk} )^2, \mbox{ and }\\
D_k = &1 + \rho_d \sum_{m=1}^M \eta_{mk} \gamma_{mk}
\beta_{mk} +  \rho_d   \sum_{i \neq k}  [ \sum_{m=1}^M  \eta_{mi}
\beta_{mk}  \lVert {\bf a}_{m,i} \rVert^2_2 \\
&+
  \tau \rho_p ( | \sum_{m=1}^M
\eta_{mi}^{1/2} \beta_{mk}  \boldsymbol{\psi}^H_k {\bf a}_{m,i} |^2+
\sum_{m=1}^M\eta_{mi} \sum_{j=1}^K \beta_{mk} \beta_{mj} |
\boldsymbol{\psi}^H_j {\bf a}_{m,i} |^2 ) ].
\end{align*}
\emph{{\bf Proof}}: See Appendix C. \qed

\emph{Remark: Orthogonal pilot scenario}
For the case when the users are assigned orthonormal pilots, it can be verified
that
{
\small
\begin{equation}
\begin{aligned}
{\rm SINR}^d_k = & \frac{ \rho_d  ( \sum_{m=1}^M  \eta_{mk}^{1/2}
\gamma_{mk} )^2} {  1 + \rho_d \sum_{m=1}^M \eta_{mk} \gamma_{mk}
\beta_{mk} +  \rho_d   \sum_{i \neq k}  [  \sum_{m=1}^M \eta_{mi}
\frac{\beta_{mk} \gamma_{mi}}{1+ \tau \rho_p \beta_{mi}} + \tau \rho_p
\sum_{m=1}^M \eta_{mi} \beta_{mk} \beta_{mi} |  \boldsymbol{\psi}_i^H {\bf
a}_{m,i} |^2  ]},\\
=& \frac{ \rho_d  ( \sum_{m=1}^M  \eta_{mk}^{1/2} \gamma_{mk} )^2} {
1 +  \rho_d \sum_{i=1}^K  \sum_{m=1}^M \eta_{mi}  \gamma_{mi} \beta_{mk}  },
\end{aligned}
\label{EqRem3D}
\end{equation}}
\normalsize
which corresponds to the SINR in (25) of ~\cite{Hien_CF}, assuming that the pilots
are orthonormal. In the case when the APs are collocated, we have $\beta_{mk}
\triangleq \beta_{k}$, $\gamma_{mk} \triangleq \gamma_k$, and the $M$ power constraints for APs are replaced by
the total power constraint, $\sum_{j=1}^K \eta_j \leq
1$. Thus, $\eta_{mk} = \eta_k / (M \gamma_{k})$ and (\ref{EqRem3D})
simplifies to

\begin{equation}
\begin{aligned}
{\rm SINR}^d_k = & \frac{ \rho_d M \eta_k} {  1 +  \rho_d \beta_{k} \sum_{i=1}^K
\eta_i   },
\end{aligned}
\label{EqRem4D}
\end{equation}
which is the downlink SINR obtained in~\cite{Hien_largeMU,yang_large}.

\section{Downlink power control}
\label{DownPower}
In the downlink, we aim to optimize the downlink power   coefficients $\eta_{mk},
\, m=1,\dots,M, \,\, k=1,\dots,K,$ so as to maximize the minimum downlink rate
of all the users. That is
\begin{equation}
\begin{aligned}
& {\rm max}_{\{ \eta_k \}} \,\, {\rm min}_{k=1,\dots,K} \,\, {\rm SINR}_k^d \\
& \text{subject to}\,\,  \sum_{k=1}^K \eta_{mk} \gamma_{mk} \leq 1,
m=1,\dots,M, \\
& \hspace{1.5cm} \eta_{mk} \geq 0,  k=1,\dots,K, \,\, m=1,\dots,M.
\end{aligned}
\label{EqDPC1}
\end{equation}
This optimization problem is quasi-concave. It can be solved by
performing a bisection search with a convex feasibility problem in each
step. More specifically, we reformulate the problem as
\begin{equation}
\begin{aligned}
 {\rm max}_{\{ \eta_k \}, t} \,\,  t  \\
 \text{subject to}\,\, & t \leq {\rm SINR}_k^u, \,\, k=1,2,\dots,K, \\
&    \sum_{k=1}^K \eta_{mk} \gamma_{mk}\leq 1, m=1,\dots,M, \\
& \eta_{mk} \geq 0,  k=1,\dots,K, \,\, m=1,\dots,M.
\end{aligned}
\label{EqDPC2}
\end{equation}
In~(\ref{EqDPC2}), we can rewrite the constraint $t \leq {\rm SINR}_k^u$ as  a
second-order cone program. Let $\zeta_{mk} = \eta_{mk}^{0.5}$. Then, $t \leq {\rm
SINR}_k^u$ is equivalent to~\cite{yang2018energy}
\begin{equation}
\begin{aligned}
{\bf b}_{k}^T \boldsymbol{\zeta} \geq \sqrt{t} \lVert   {\bf C}_{k}
\boldsymbol{\zeta} \rVert_2,
\end{aligned}
\label{EqDPC3}
\end{equation}
where
\begin{align*}
\boldsymbol{\zeta}&= [1,\zeta_{11},\zeta_{21},\dots,\zeta_{M1},
\dots, \zeta_{1K},\zeta_{2K},\dots,\zeta_{MK} ]^T \in
\mathbb{R}_{0^+}^{(MK+1) \times 1}, \\
{\bf b}_k &= [  0,0,\dots,0,
\gamma_{1k},\gamma_{2k},\dots,\gamma_{Mk},0,\dots,0 ]^T   \in
{\mathbb{R}_{0+}^{(MK +1) \times 1}}.
\end{align*}
Furthermore, ${\bf C}_k = [ {\bf
F}_k; {\bf P}_k ] \in \mathbb{C}^{(MK+K+1) \times (MK+1)}$, where ${\bf
F}_k \in \mathbb{R}^{(MK+1) \times (MK+1)} $ and ${\bf P}_k \in \mathbb{C}^{K
\times (MK+1)}$. Here, ${\bf F}_k$ is given by
\begin{equation}
{\bf F}_k = {\rm diag} ( 1/\sqrt{\rho_d}, \sqrt{f^{k}_{1,1}}, \dots,
\sqrt{f^{k}_{M,1}}, \dots, \sqrt{\beta_{1k} \gamma_{1k}}, \dots,
\sqrt{\beta_{Mk} \gamma_{Mk}}, \dots, \sqrt{f^{k}_{1,K}}, \dots,
\sqrt{f^{k}_{M,K}} ),
\label{EqDPC3A}
\end{equation}
where for $i \neq k, f^{k}_{m,i} = \beta_{mk} \lVert {\bf a}_{m,i} \rVert_2^2 +
\tau \rho_p \sum_{j=1}^K \beta_{mj}  \beta_{mk} | \boldsymbol{\psi}_j^H
{\bf a}_{m,i} |^2 $. Also, ${\bf P}_k = [  {\bf p}_1^{k}; {\bf
p}_2^{k}; \dots; {\bf p}_{K}^{k} ]$. The $i$-th row of ${\bf P}_k$, ${\bf
p}_i^{k}$, is given by
\begin{equation}
{\bf p}_i^{k} = \begin{cases}
 [ 0,\dots,0,   \sqrt{\tau \rho_p} \beta_{1k} \boldsymbol{\psi}_k^H {\bf
a}_{1,i}, \dots, \sqrt{\tau \rho_p} \beta_{Mk} \boldsymbol{\psi}_k^H {\bf
a}_{M,i}  , 0,\dots, 0  ], \quad \text{if } i \neq k, \\
[0,0,\dots,0], \quad \text{if } i=k.
\end{cases}
\label{EqDPC3B}
\end{equation}
Thus, the non-zero elements of ${\bf p}_i^{k}$ span the columns $(i-1)M+2$ to
$iM +1$ of ${\bf p}_i^{k}, i \neq k$. Since ${\bf C}_k$ in~(\ref{EqDPC3}) is complex valued, we rewrite the constraint
as
\begin{equation}
\begin{aligned}
{\bf b}_{k}^T \boldsymbol{\zeta} \geq \sqrt{t} \lVert  [
\begin{array}{c}
{\rm Re}({\bf C}_{k})\\
{\rm Im}({\bf C}_{k})
\end{array} ] \boldsymbol{\zeta} \rVert_2, \quad k=1,\dots,K.
\end{aligned}
\label{EqDPC4}
\end{equation}
Similarly, we can write the constraint $\sum_{k=1}^K \eta_{mk} \gamma_{mk} \leq
1$ as
\begin{equation}
\begin{aligned}
\lVert {\bf Z}_m \boldsymbol{\zeta} \rVert_2  \leq 1, \,\,
m=1,2,\dots,M,
\end{aligned}
\label{EqDPC5}
\end{equation}
where ${\bf Z}_m = {\rm diag} (0, {\bf z}_{1,m}, {\bf z}_{2,m}, \dots, {\bf
z}_{K,m} ) \in \mathbb{R}_{0^+}^{(MK+1) \times (MK+1)}$ with ${\bf
z}_{k,m} = \sqrt{\gamma_{mk}} {\bf e}_m$, ${\bf e}_m$ being the $m$-th column of
the identity matrix ${\bf I}_M$. For a given $t$, the CVX package~\cite{cvx} is
used to test whether the constraints~(\ref{EqDPC4}) and~(\ref{EqDPC5}) are
feasible. Then, bisection search gives us the maximum $t$ corresponding to the
maximum SINR attainable $\forall k$.

\section{Small-Cell Systems}
\label{SmallCell}
Small-cell systems have been suggested as possible solutions to support the data traffic of next generation wireless systems. They consist of a dense deployment of low-power APs, and are characterized by small path loss and reduced transmit powers~\cite{liu2014energy,liu2014comparative}. Here, we briefly summarize the description of small-cells as provided in~\cite{Hien_CF} and~\cite{nayebi2017precoding}.

The $M$ APs serve $K$ users in the same coverage area as the cell-free system. Each AP serves only one user at a time, usually the user with the strongest received signal. Since each user is served by at most one AP, the channel does not harden and the users as well as the APs must estimate their effective channels. Hence, both uplink and downlink training are needed.

As in cell-free systems, in the uplink, the users send their pilots to the APs and the APs estimate the channels. The channel estimates are then used to decode the data symbols sent by the users. Let the AP selected for the $k$-th user be $m_k$ and the power   coefficient of the $k$-th user be $\eta_k^{\rm sc}$. Let $\tau_u$ be the length of the uplink pilots and $\rho_{u,p}$ be the transmit power per uplink pilot symbol. Then, the uplink achievable rate for the $k$-th user is given in terms of the exponential integral function (Ei) by~\cite{Hien_CF}:
\begin{align}\label{eq:R_sc}
R_{\rm sc} = -({\rm log}_2 e) e^{1/ \omega_{m_k k}} {\rm Ei} ( - \frac{1}{\omega_{m_k k}}  ),
\end{align}
where
\begin{align*}
&\omega_{m_k k} = \frac{\rho_u \eta_k^{\rm sc}  \bar{\omega}_{m_k k} }{ \rho_u \eta_k^{\rm sc} ( \beta_{m_k k} - \bar{\omega}_{m_k k}) + \rho_u \sum_{k^\prime \neq k} \eta_{k^\prime}^{\rm sc} \beta_{m_k k^\prime} + 1},\\
& \bar{\omega}_{m_k k} = \frac{\rho_{u,p} \tau_u \beta_{m_k k}^2 }{\rho_{u,p} \tau_u \sum_{k^\prime = 1}^K \beta_{m_k k} | \boldsymbol{\psi}_k^H \boldsymbol{\psi}_{k^\prime} |^2 +1 }.
\end{align*}
Similarly, in the downlink, the power   coefficient for the symbol
transmitted by the $m_k$-th AP to the $k$-th user is $\alpha_k^{\rm sc}$. Let the length of the downlink pilots be $\tau_d$ and $\rho_{d,p}$ be the transmit power per downlink pilot symbol. The
achievable rate for the $k$-th user is given by~\cite{Hien_CF}:
\begin{align*}
R_{\rm sc} = -({\rm log}_2 e) e^{1/ \mu_{m_k k}} {\rm Ei} ( -
\frac{1}{\mu_{m_k k}}  ),
\end{align*}
where
\begin{align*}
&\mu_{m_k k} = \frac{\rho_d \alpha_k^{\rm sc}  \bar{\mu}_{m_k k} }{ \rho_d
\alpha_k^{\rm sc} ( \beta_{m_k k} - \bar{\mu}_{m_k k}) + \rho_d \sum_{k^\prime
\neq k} \alpha_{k^\prime}^{\rm sc} \beta_{m_{k^{\prime}} k} + 1},\\
& \bar{\mu}_{m_k k} = \frac{\rho_{d,p} \tau_d \beta_{m_k k}^2 }{\rho_{d,p} \tau_d
\sum_{k^\prime = 1}^K \beta_{m_{k^{\prime}} k} | \boldsymbol{\psi}_k^H
\boldsymbol{\psi}_{k^\prime} |^2 +1 }.
\end{align*}
The equivalent max-min power control problems are quasi-linear programs that
can be solved by bisection.
\section{Simulation Results}
\label{Sims}
We compare the performance of cell-free Massive MIMO system with LMMSE estimation with
the performance  of the system  with suboptimal channel estimation considered in~\cite{Hien_CF} and with
the performance of small-cell systems. For each user in the small-cell system, it is
assumed that the AP with the strongest received signal is selected. As a
performance measure, we use the per-user throughput which takes into
account the estimation overhead. User detection algorithm proposed in \cite{liu2018massive} requires assigning a unique pilot
to each user. In our simulations we use random pilots, that is random vectors uniformly distributed over the surface of complex unit sphere in $\mathbb{C}^\tau$. Such pilots are almost surely distinct for all $N$ users.

\subsection{Simulation Settings}
We use the path loss and shadow fading correlation model as in~\cite{tang2001mobile,Hien_CF}. The large scale fading coefficients $\beta_{mk}$
are modeled as
\begin{equation*}
\beta_{mk} = {\rm PL}_{mk} 10^{\frac{\sigma_{\rm sh} z_{mk}}{10}},
\label{EqSim1}
\end{equation*}
where ${\rm PL}_{mk} $ is the path loss and $10^{\frac{\sigma_{\rm sh}
z_{mk}}{10}}$ is the shadow fading with standard deviation $\sigma_{\rm sh}$ and
$z_{mk} \sim \mathcal{N}(0,1)$. The path loss is defined by the three-slope
model:
$$
{\rm PL}_{mk}=\left\{
\begin{array}{ll}
-L-34\log_{10}(d_{mk}) & d_{mk} > d_1,\\
-L-15\log_{10}(d_{1})-20\log_{10}(d_{mk}), & d_0<d_{mk} \le d_1,\\
-L-15\log_{10}(d_{1})-20\log_{10}(d_{0}) & d_{mk} \le d_0,
\end{array}\right.
$$
where
$$
L\triangleq 46.3 + 33.9 \log_{10}(f)
-13.82 \log_{10}(h_{AP})
-(1.1 \log_{10}( f ) -0.7 )h_u + (1.56 \log_{10}( f ) -0.8),
$$
and $f$ is the carrier frequency (in MHz), $h_{AP}$ is the AP
antenna height (in m), and $h_u$ denotes the user antenna height
(in m).

Further, the shadow fading is defined by the model with two components:
$$
z_{mk}=\sqrt{\delta}a_m+\sqrt{1-\delta}b_k, m=1,\ldots,M,~k=1,\ldots,M,
$$
where $a_m,b_k\sim {\cal N}(0,1)$ and $\delta$ and $0\le \delta \le 1$ is a parameter.
The covariances of $a_m$ and $b_k$ are given by:
$$
\mathbb{E}[{ a_m \, a_{m'} }] = 2^{-d_a(m,m') \over d_{decorr}},~
\mathbb{E}[{ b_k \,b_{k'}}] = 2^{-d_u(k,k') \over d_{decorr}},
$$
where $d_a(m,m')$ and $d_u(k,k')$ are the distances between APs and users respectively, and
$d_{decorr}$ is a decorrelation distance, which depends on the environment.

The noise power is computed as follows:
\begin{equation*}
\text{noise power} = \text{bandwidth} \times k_{\rm B} \times T_0 \times \text{noise figure},
\label{EqSim2}
 \end{equation*}
where $k_{\rm B}$ is the Boltzmann constant ($1.381 \times 10^{-23}$ J/K),
$T_0$ is the noise temperature ($290$~Kelvin $\approx 17$ Celsius). The noise figure is set to $9$ dB and
the bandwidth is $20$~MHz. We consider a square $D\times D$ area of the network
wrapped around to edges to avoid boundary effects. As a performance measure, we
consider per-user net throughput. For the cell-free case, the throughput is
defined as
\begin{equation*}
\text{S}_k^{u,{\rm cf}} = B \frac{1-\tau/\tau_c}{2} R^{u,{\rm cf}}_k,
\label{EqSim3}
\end{equation*}
and for the small-cell system, it is defined as
\begin{equation*}
\text{S}_k^{u,{\rm sc}} = B \frac{1-(\tau_u + \tau_d)/\tau_c}{2} R^{u,{\rm sc}}_k,
\label{EqSim4}
\end{equation*}
 where $R^{u,{\rm cf}}_k$ and $R^{u,{\rm sc}}_k$ are the achievable uplink
rates  given by (\ref{R_k^u,cf}) and (\ref{eq:R_sc}) respectively, $B$ is the spectral bandwidth, and
$\tau_c$ is the coherence interval measured in OFDM symbols. In the small-cell systems, we spend $\tau_u$ and $\tau_d$ samples for
uplink and downlink training respectively. In all our simulations below we use $\tau_u=\tau_d=\tau$.

In our simulations we use the settings defined in the following table.

\begin{center}\label{tab:settings}
\begin{tabular}{|c|c|}
\hline
Parameter & Value \\
\hline
Carrier Frequency & $1.9$~GHz \\
\hline
AP Antenna Height & $15$~m\\
\hline
User Antenna Height & $1.65$~m \\
\hline
$\sigma_{sh}$ & $8$~dB \\
\hline
$d_1$, $d_0$ & $50$, $10$~m\\
\hline
$d_{decorr}$ & $20$~m \\
\hline
\end{tabular}
\end{center}

\subsection{Results and Discussions}
In this section we compare the performance of cell-free systems with LMMSE and suboptimal channel estimations, and the performance of small-cells systems. We consider the scenario of users with transmit power of $20$~mW,  which is relevant for  IoT networks, as well as the scenario of users with transmit power of $200$~mW, which is a standard assumption for mobile communication networks. In our simulations we consider the max-min and target power controls, as well as uniform maximum transmit powers, i.e., no power control.

To plot the CDF of the per-user
throughput, we use $500$ random realizations of the channel and pilots. For the
small-cell case we assume  $\rho_{u,p} = \rho_{d,p} = \rho_{u}$.

\begin{figure}[h!]%
    \centering
  {{\includegraphics[width=3.5in]
{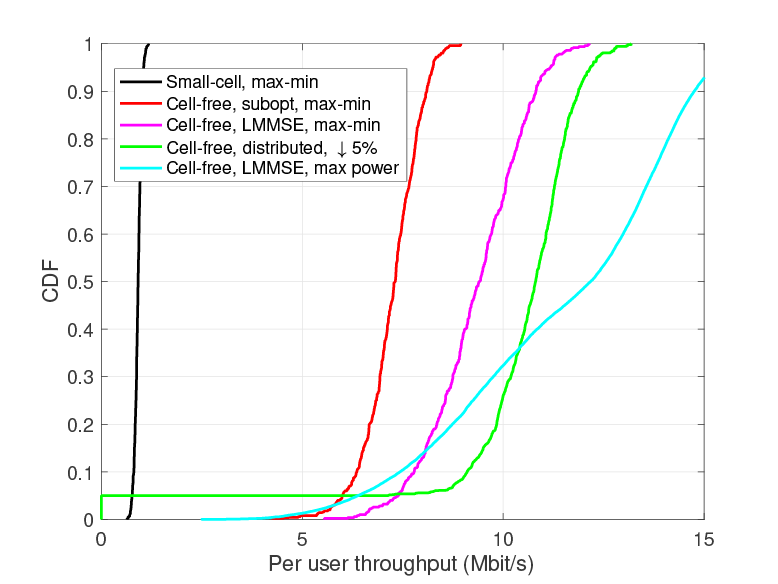}}
    \caption{CDF of uplink user throughputs with uncorrelated shadow fading. $M=128, K=40, \tau =60$, cell size $D=100$~m and $\rho_u=\rho_p = 20$~mW. }
\label{fig:3}}
    \end{figure}

\emph{Uplink Experiment 1}: In the IoT scenario, we assume that the pilot length is
greater than the number of active users. We consider $M=128$, $K=40$, $\tau =
60$, $D=100$~m, and uncorrelated shadow fading. The transmit power is $\rho_u =
\rho_p = 20$~mW, which is one magnitude lower than the power  of a typical mobile
phone. The CDFs of the SINRs achieved by all users under different power control
schemes are plotted in Fig.~\ref{fig:3}. The sign $\downarrow$ indicates that
$5\%$ of the users are dropped. More specifically, the target SINR is adjusted
such that $5\%$ of the users cannot achieve it. One can see that all cell-free
systems gain very significantly over the small-cell system. Our cell-free
system with LMMSE estimation and the max-min power control gives 30\% gain over
the system with suboptimal channel estimation in terms of $95$\% outage rate
(the lowest rate among the best 95\% of the users). Our cell-free system with
distributed power control and elimination of $5\%$ of the users gives about
$40\%$ in term of the median throughput.
Additional simulations (not shown here) indicate that not much can be gained by
selecting the AP with the maximum $\beta_{mk}$ for small-cell systems.

\begin{figure}[h!]%
    \centering
 {{\includegraphics[width=3.5in]
{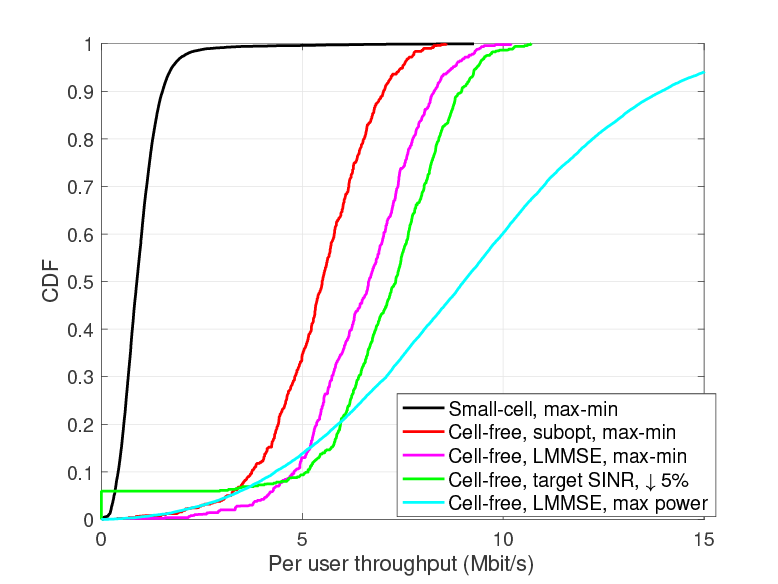} }
    \caption{CDF of uplink user throughputs with correlated shadow fading. $M=128, K=40, \tau =60$, cell size $D=500$~m and $\rho_u=\rho_p = 20$~mW. }
\label{fig:4}}
    \end{figure}

A similar
effect is observed in Fig.~\ref{fig:4} with correlated shadow fading and
$D=500$~m. Based on the results in~\cite{chen2018channel}, cell-free
systems exhibit slower channel hardening, capacity lower bounds that rely on channel hardening
(such as our ``use-then-forget" achievable rate bound) can be quite loose.
Therefore, our performance estimates for the 50\% likely throughput are
conservative.

\begin{figure}[h!]%
    \centering
 {{\includegraphics[width=3.5in]
{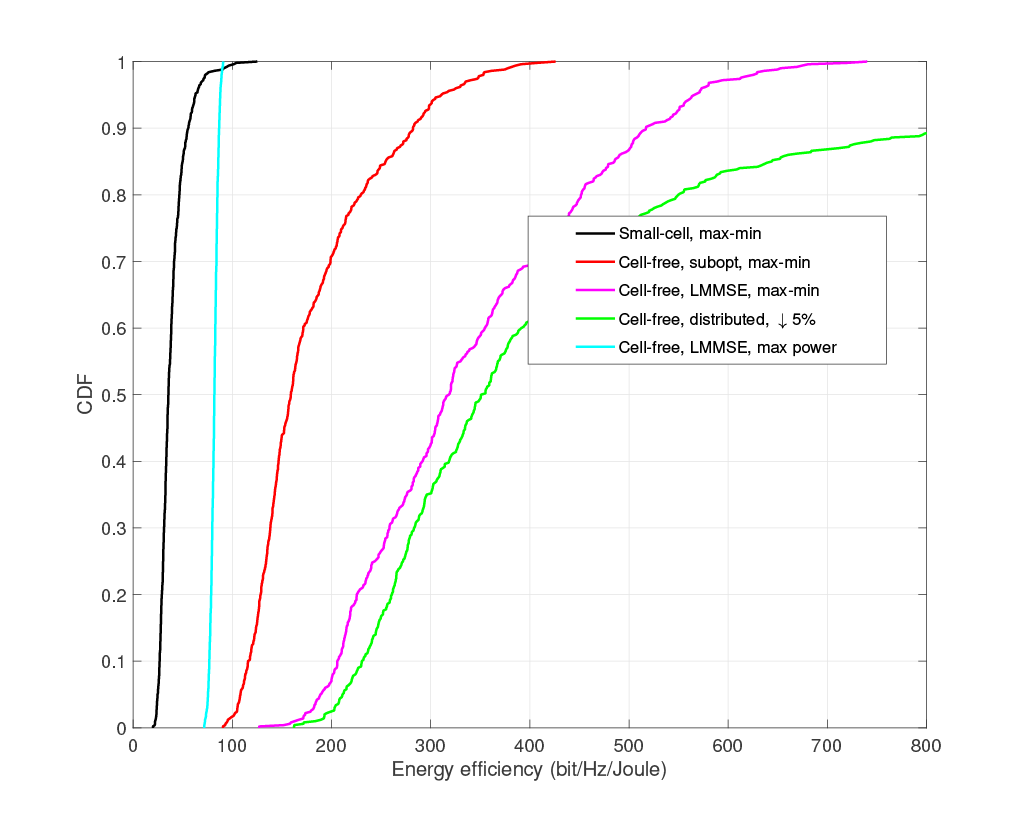} }
    \caption{CDF of the energy efficiency of uncorrelated users. $M=128, K=40, \tau =60$, cell size $D=500$~m and $\rho_u=\rho_p = 20$~mW. }
\label{fig:EEuncorr}}
    \end{figure}

\begin{figure}[h!]%
    \centering
 {{\includegraphics[width=3.5in]
{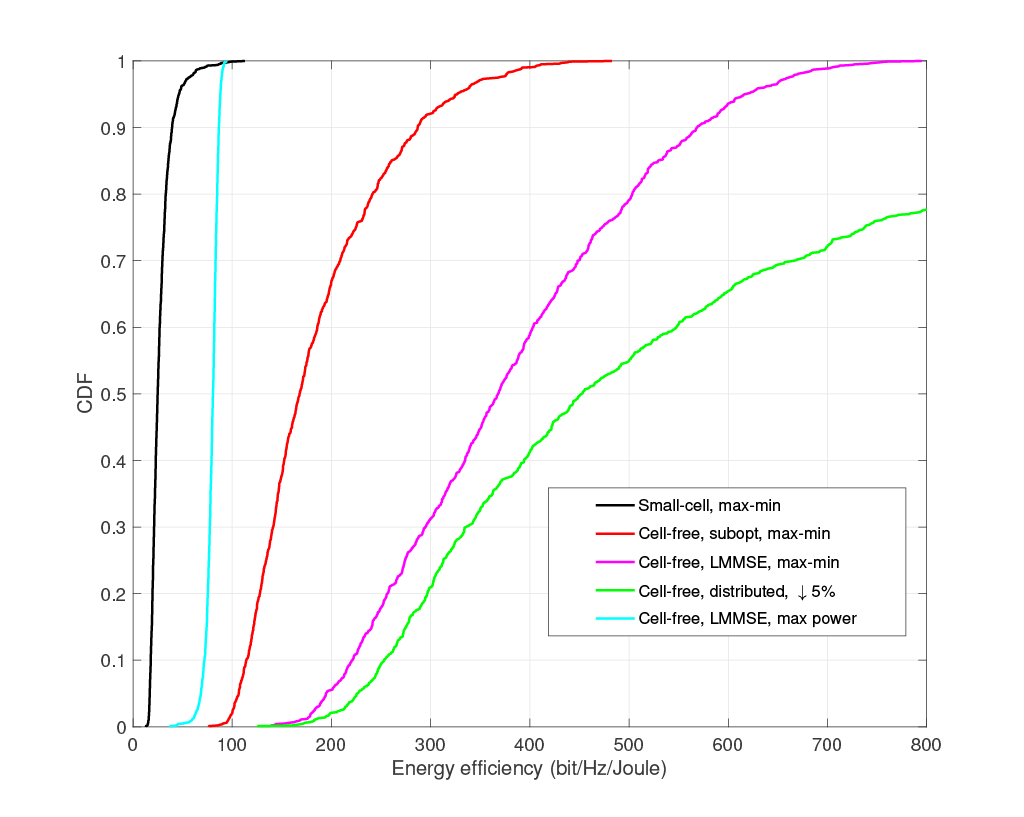} }
    \caption{CDF of the energy efficiency of correlated users. $M=128, K=40, \tau =60$, cell size $D=500$~m and $\rho_u=\rho_p = 20$~mW. }
\label{fig:EEcorr}}
    \end{figure}

In Fig.~\ref{fig:EEuncorr} and Fig.~\ref{fig:EEcorr} we present results on the energy efficiency obtained with small-cell and cell-free systems with different power control algorithms. We observe that in terms of the energy efficiency power control algorithms and LMMSE channel estimation give even larger gains, than in terms of  uplink data rates. In particular, systems with LMSSE channel estimation and max-min power control and and distributed power control gain 150\% and 180\% respectively over maximal  power transmission. They also gain 74\% and 95\% respectively compared with the systems with suboptimal channel estimation and max-min power control. Similar results hold for correlated users, Fig.~\ref{fig:EEcorr}.

\begin{figure}[h!]%
    \centering
{{\includegraphics[width=3.5in]
{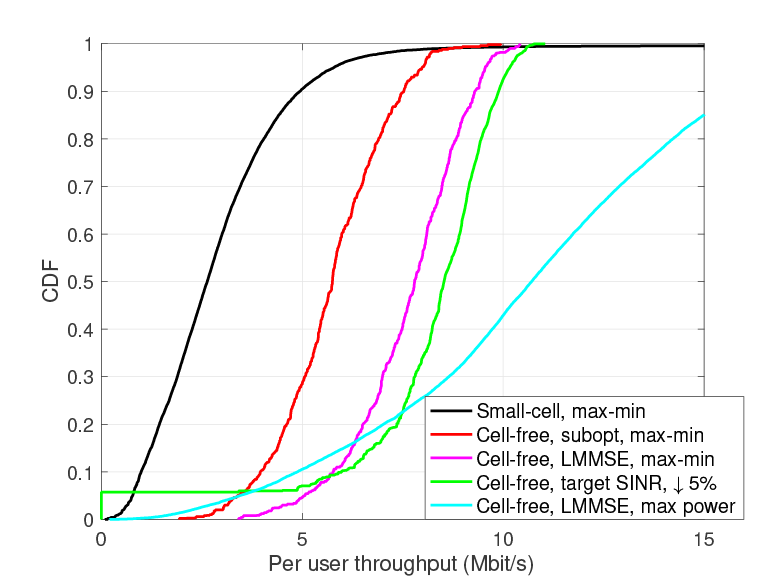}}
    \caption{CDF of uplink user throughputs with uncorrelated shadow fading.
$K=40, \tau =20$, cell size $D=1$~km and $\rho_u=\rho_p = 200$~mW. }
\label{fig:1}}
    \end{figure}

  \emph{Uplink Experiment 2}:  Here, we assume a mobile phone communication scenario,
rather than IoT scenario. We assume $D=1$~km and the transmit power $\rho_u =
\rho_p = 200$~mW, and uncorrelated shadow fading. Results are presented in
Fig.~\ref{fig:1}. It can be seen that the cell-free systems outperform the
small-cell system in both the median and the $95\%$-outage rate. In particular,
in terms of the $95\%$-outage rate, the cell-free system with suboptimal
channel
estimation gives a $5$-fold improvement. The cell-free system with LMMSE
channel estimation and max-min power control gives a 50\% improvement in terms
of $95\%$-outage rate. The system with the distributed power control and $5\%$
users dropping, that is 2 users in our settings, the median throughput can be
further improved by about $10\%$. Note also that the $95\%$-outage rate with
optimal channel estimation is about $50\%$ greater than in the case of no power
control (max power).

\begin{figure}[h!]%
    \centering
  {{\includegraphics[width=3.5in]
{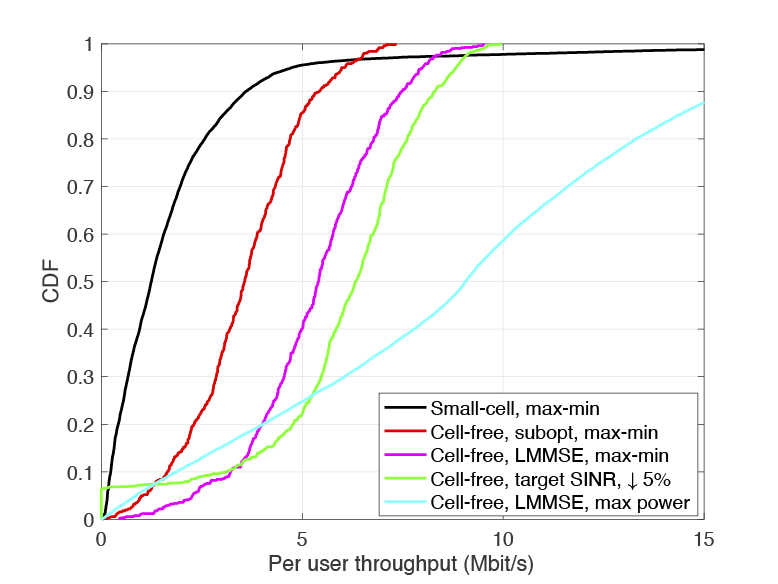} }
    \caption{CDF of uplink user throughputs with correlated shadow fading. $K=40, \tau =20$, cell size $D=1$~km and $\rho_u=\rho_p = 200$~mW. }
\label{fig:2}}
    \end{figure}

Fig.~\ref{fig:2} presents results for the scenario with correlated shadow
fading. One can see that correlated shadowing results in significant reduction
of  the throughputs. Still, the $95\%$-outage rate likely throughput of the
cell-free system with LMMSE channel estimation and distributed power control is
around 10 times higher than that of small-cell. It is important to note that the
$95\%$-outage rate of cell-free system with the optimal channel estimation is
twice that of the system with suboptimal channel estimation. These energy efficiency gains can be
crucially important for IoT systems.

\begin{figure}[h]%
    \centering
{{\includegraphics[width=3.5in]
{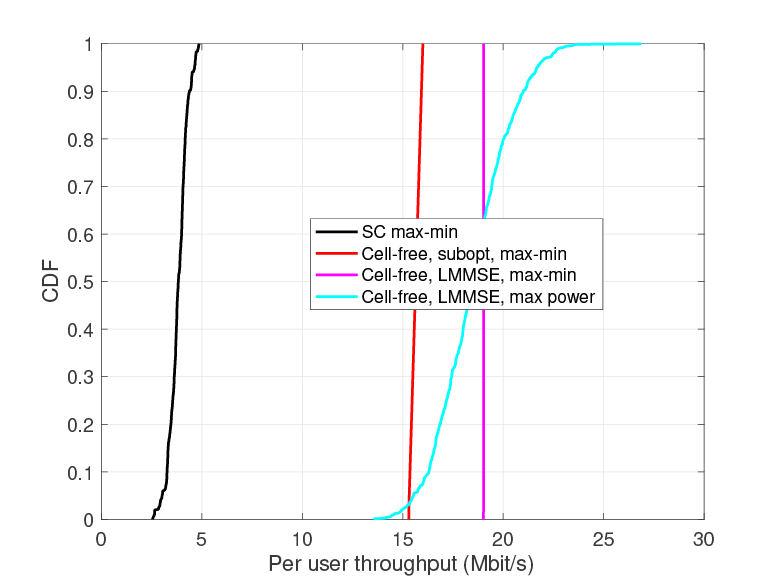}}
    \caption{CDF of downlink user throughputs with uncorrelated shadow fading.
$K=20, \tau =40$, cell size $D=100$~m and $\rho_u=\rho_p = 20$~mW. }
\label{fig:5}}
    \end{figure}

 \emph{Downlink Experiment 1}:  In this scenario, we consider the downlink throughputs with $M=128, \,K = 20, \,\tau = 40$. The cell size $D=100$~m, the shadow fading is uncorrelated and $\rho_u = \rho_d = 20$~mW. Fig.~\ref{fig:5} shows the per-user downlink throughput. The cell-free throughputs are about three times the throughput of the small-cell system. With LMMSE channel estimation, the 95\% outage-rate of the cell-free system is about 25\% higher than with suboptimal channel estimation.

\begin{figure}[h]%
    \centering
  {{\includegraphics[width=3.5in]
{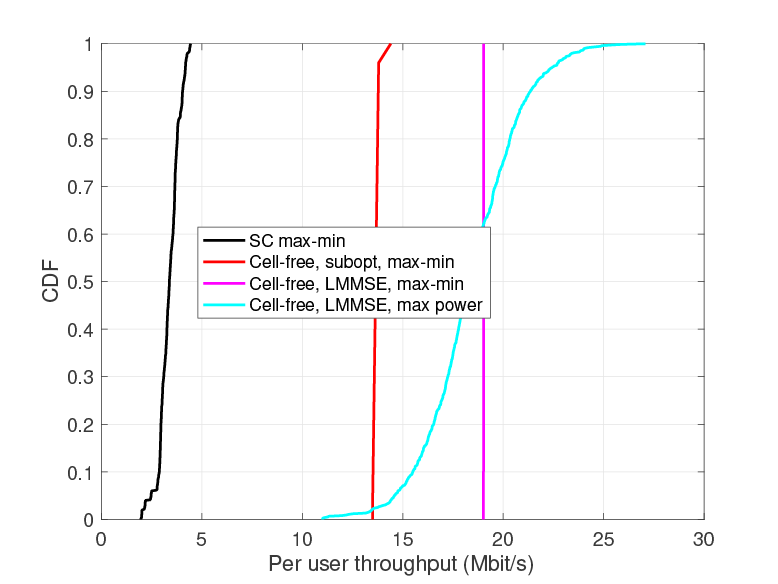} }
    \caption{CDF of downlink user throughputs with correlated shadow fading. $K=20, \tau =40$, cell size $D=100$~m and $\rho_u=\rho_p = 20$~mW. }
\label{fig:6}}
    \end{figure}

 A similar result is seen for correlated shadow fading in Fig.~\ref{fig:6}. Here, the 95\% outage-rate with optimal channel estimation is about 40\% higher than with suboptimal channel estimation.

\begin{figure}[h]%
    \centering
  {{\includegraphics[width=3.5in]
{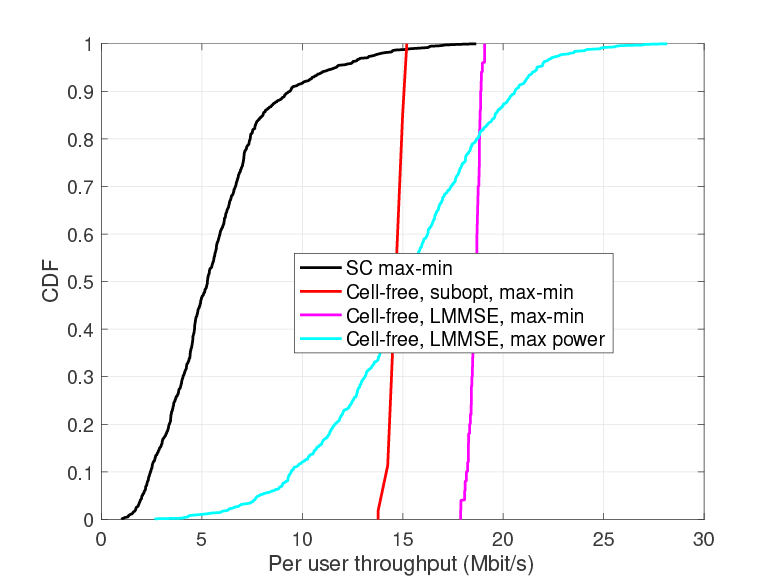} }
    \caption{CDF of downlink user throughputs with correlated shadow fading. $K=20, \tau =10$, cell size $D=1$~km and $\rho_u=\rho_p = 200$~mW. }
\label{fig:7}}
    \end{figure}

  \emph{Downlink Experiment 2}:  In this case, we assume a mobile phone communication scenario with $D = 1$km, $\rho_u=\rho_d = 200$~mW and uncorrelated shadow fading. The results are shown in Fig.~\ref{fig:7}. Both in the median and the 95\% outage-rate, the cell-free systems outperform the small-cell system.  The throughput with LMMSE channel estimation is about 28\% higher than the cell-free system with suboptimal channel estimation.

\section{Conclusion}
We considered cell-free massive MIMO for IoT with optimal (LMMSE) channel estimation and random uplink pilots. The goal for studying such system was to considerably improve the spectral efficiency of both uplink and downlink transmissions compared with cell-free massive MIMO systems with sub-optimal channel estimation, and  small-cell systems. The second goal was
developing transmit power control algorithm that allow infrequent power adaptation, and provide additional improvement of the system performance (spectral efficiency). High energy efficiency is  especially important for IoT because the things are
operating under low power conditions. Centralized max-min SINR and distributed
target SINR power control algorithms are utilized. Simulation results with max-min SINR power optimization show more than ten-fold improvement over conventional IoT systems based on
small-cell architecture. A 40\% improvement over cell-free systems with
suboptimal channel estimation is achieved in terms of uplink data rates and 95\% in terms of energy efficiency. Similarly, in the downlink, a three-fold improvement over small-cell systems was observed along with a maximum of 40\% improvement in the downlink throughput over cell-free systems using suboptimal channel estimation.

\section*{Appendix A}\label{AppA}
{
For $T_1$ we have
\begin{equation}\label{Eq9}
T_1 =  \sqrt{\rho_u \eta_k} \sum_{m=1}^M \mathbb{E}[ (
\hat{g}_{mk} +  \tilde{g}_{mk} )  \hat{g}^*_{mk} ]
=  \sqrt{\rho_u \eta_k} \sum_{m=1}^M  \mathbb{E}[ |\hat{g}_{mk}|^2 ]
=\sqrt{\rho_u \eta_k} \sum_{m=1}^M \gamma_{mk}.
\end{equation}
It is easy  find the variance of $T_4$ is
 $$
\mathbb{E}[|{\rm T_4}|^2] =  \sum_{m=1}^M \gamma_{mk}.
$$
Finding the variances of $T_2$ and $T_3$ require longer calculations, which are presented below.

\begin{equation}
\begin{aligned}
\mathbb{E}[|T_2|^2] =& \rho_u \eta_k \sum_{m=1}^M
\sum_{n=1}^M \mathbb{E} [ ( \hat{g}^*_{mk} g_{mk} -
\mathbb{E}[\hat{g}^*_{mk} g_{mk}]  ) ( \hat{g}_{nk} g^*_{nk}
-\mathbb{E}[\hat{g}_{nk} g^*_{nk}] )  ] \\
=& \rho_u \eta_k \sum_{m=1}^M \sum_{n=1}^M [  \mathbb{E} [  \hat{g}^*_{mk}
g_{mk} \hat{g}_{nk} g^*_{nk} ] -   \mathbb{E} [  \hat{g}^*_{mk} g_{mk}]
\gamma_{nk} -  \mathbb{E} [  \hat{g}_{nk} g^*_{nk} ] \gamma_{mk} ] +
\rho_u \eta_k \sum_{m=1}^M \sum_{n=1}^M \gamma_{mk} \gamma_{nk}\\
=& \rho_u \eta_k \sum_{m=1}^M \sum_{n=1}^M \mathbb{E} [  \hat{g}^*_{mk}
(\hat{g}_{mk} + \tilde{g}_{mk}) \hat{g}_{nk} (\hat{g}^*_{nk} + \tilde{g}^*_{nk})
] - \rho_u \eta_k \sum_{m=1}^M \sum_{n=1}^M\gamma_{mk} \gamma_{nk}    \\
=& \rho_u \eta_k \sum_{m=1}^M \sum_{n=1}^M \{ \mathbb{E} [ |
\hat{g}_{mk} |^2   |  \hat{g}_{nk} |^2 ] + \mathbb{E}[ |
\hat{g}_{mk} |^2  \hat{g}_{nk} \tilde{g}^*_{nk} ] +\mathbb{E}[ |
\hat{g}_{nk} |^2  \hat{g}^*_{mk} \tilde{g}_{mk}  ]  + . \\
& \hspace{5cm} . \underbrace{\mathbb{E}[   \hat{g}^*_{mk} \tilde{g}_{mk}
\hat{g}_{nk} \tilde{g}^*_{nk}   ]}_{=0\, {\rm if } \,m \neq n} \}- \rho_u
\eta_k \sum_{m=1}^M \sum_{n=1}^M\gamma_{mk} \gamma_{nk}  \\
&\hspace{2cm}. \sum_{n=1}^M \{  \mathbb{E}[ |  \hat{g}_{mk}
|^2 \hat{g}_{nk} \tilde{g}^*_{nk} ] +  \mathbb{E}[ |  \hat{g}_{nk}
|^2 \tilde{g}_{mk} \hat{g}^*_{mk} ] \} ] - \rho_u \eta_k
\sum_{m=1}^M  \sum_{n=1}^M \gamma_{mk} \gamma_{nk}  \\
=& \rho_u \eta_k \sum_{m=1}^M [  2 \gamma^2_{mk}  + \gamma_{mk}
(\beta_{mk}- \gamma_{mk}) + \underbrace{ \mathbb{E}[| \hat{g}_{mk}
|^2 \hat{g}_{mk} \tilde{g}^*_{mk} ] + \mathbb{E}[| \hat{g}_{mk}
|^2 \hat{g}^*_{mk} \tilde{g}_{mk} ]}_{=0} + . \\
& . \hspace{1cm} \sum_{n \neq m} \{ \gamma_{mk} \gamma_{nk} +
\underbrace{ \mathbb{E}[ | \hat{g}_{mk} |^2  \hat{g}_{nk}
\tilde{g}^*_{nk} ] + \mathbb{E}[| \hat{g}_{nk} |^2  \hat{g}^*_{mk}
\tilde{g}_{mk} ]}_{=0} \}   - \sum_{n=1}^M \gamma_{mk} \gamma_{nk}
]\\
=& \rho_u \eta_k \sum_{m=1}^M [ \beta_{mk}\gamma_{mk} +  \sum_{n=1}^M
\gamma_{mk} \gamma_{nk} ] - \rho_u \eta_k \sum_{m=1}^M \sum_{n=1}^M
\gamma_{mk} \gamma_{nk} = \rho_u \eta_k \sum_{m=1}^M \gamma_{mk} \beta_{mk}.
\end{aligned}
\label{Eq10}
\end{equation}

Further,
$$
\mathbb{E}[|T_3|^2] = \sum_{i \neq k} \mathbb{E}[
|v_i|^2 ].
$$
We will consider individual terms of this sum:
\begin{align}
\mathbb{E}[ |v_i|^2 ]  = & \rho_u \eta_i \mathbb{E} [|
\sum_{m=1}^M g_{mi} \hat{g}^*_{mk} |^2 ] \nonumber  \\
=&  \rho_u \eta_i \mathbb{E} [  | \sum_{m=1}^M g_{mi} (  {\bf
a}^H_{m,k}[ \sqrt{\tau \rho_p} \boldsymbol{\Psi} {\bf g}_m + {\bf w}_m
]  )^* |^2    ] \nonumber \\
=& \rho_u \eta_i \mathbb{E} [ | \sum_{m=1}^M g_{mi} \sum_{j=1}^{K}
\sqrt{\tau \rho_p} g^*_{mj} \boldsymbol{\psi}^H_j {\bf a}_{m,k} + \sum_{m=1}^M
g_{mi} {\bf w}^H_m {\bf a}_{m,k} |^2  ]  \nonumber \\
=&  \rho_u \eta_i \tau \rho_p \mathbb{E} [ | \sum_{m=1}^M g_{mi}
\sum_{j=1}^{K}  g^*_{mj} \boldsymbol{\psi}^H_j {\bf a}_{m,k} |^2] +
\rho_u \eta_i   \mathbb{E} [ |  \sum_{m=1}^M g_{mi} {\bf w}^H_m {\bf
a}_{m,k} |^2 ] \nonumber \\
=& \rho_u \eta_i \sum_{m=1}^M \beta_{mi}  \lVert {\bf a}_{m,k} \rVert^2_2 +
\rho_u \eta_i \tau \rho_p\underbrace{ \sum_{m=1}^M \sum_{n=1}^M \sum_{j=1}^K
\sum_{l=1}^K  \mathbb{E} [    g_{mi}  g^*_{mj} g^*_{ni}  g_{nl}
\boldsymbol{\psi}^H_j {\bf a}_{m,k}   {\bf a}^H_{n,k} \boldsymbol{\psi}_l
]}_{\mathcal{T}_i}.\label{Eq11}
\end{align}

For finding $\mathcal{T}_i$ we first note that

\begin{equation}
\begin{aligned}
\mathbb{E} [    g_{mi}  g^*_{mj} g^*_{ni}  g_{nl} ] = & \begin{cases}
{\rm if}\, m=n \begin{cases}
\mathbb{E}[ |  g_{mi} |^4 ] = 2 \beta^2_{mi}, \quad {\rm
if } \, j=l=i, \\
\mathbb{E}[ |  g_{mi} |^2  |  g_{mj} |^2 ]=
\beta_{mi} \beta_{mj}, \quad {\rm if } \, j=l \neq i, \\
0, \quad {\rm otherwise.}\\
\end{cases}\\
{\rm if}\, m \neq n \begin{cases}
\mathbb{E} [   | g_{mi} |^2 | g_{ni}|^2 ] =
\beta_{mi} \beta_{ni}, \quad {\rm if } \, j=l=i, \\
0, \quad {\rm otherwise.}\\
\end{cases}
\end{cases}
\end{aligned}
\label{Eq12}
\end{equation}

Thus,

\begin{equation}
\begin{aligned}
\mathcal{T}_i =  \sum_{m=1}^M ( 2 \beta^2_{mi} |\boldsymbol{\psi}^H_i
{\bf a}_{m,k} |^2 +   \sum_{j \neq i}^K \beta_{mi} \beta_{mj}
|\boldsymbol{\psi}^H_j {\bf a}_{m,k} |^2 + \sum_{n \neq m}
\beta_{mi} \beta_{ni} \boldsymbol{\psi}^H_i {\bf a}_{m,k}   {\bf a}^H_{n,k}
\boldsymbol{\psi}_i   ).
\end{aligned}
\label{Eq13}
\end{equation}

Using these results, we obtain
\begin{equation}
\begin{aligned}
\mathbb{E}[|T_3|^2] =& \rho_u \eta_i \sum_{m=1}^M \beta_{mi}
 {\bf a}^H_{m,k}  {\bf a}_{m,k} +  \rho_u \eta_i \tau \rho_p \sum_{m=1}^M (
 \sum_{n=1}^M \beta_{mi} \beta_{ni} \boldsymbol{\psi}^H_i {\bf a}_{m,k} {\bf
a}^H_{n,k} \boldsymbol{\psi}_i + \sum_{j=1}^K \beta_{mi} \beta_{mj} |
\boldsymbol{\psi}^H_j {\bf a}_{m,k} |^2 ).
\end{aligned}
\label{Eq14}
\end{equation}

}
\section*{Appendix B}
\label{AppB}

To prove the theorem, we have to show that the functions $I_k(\boldsymbol{\eta})
= \frac{\eta_k}{{\rm SINR}_k^u}$ and $1/I_k(\boldsymbol{\eta})$ are two-sided
scalable~\cite{rasti2011distributed,adhikary2017uplink}.

A function $I(\boldsymbol{\eta})$ is two-sided scalable if for any  $\alpha >
1$, and vectors $\boldsymbol{\eta}_1$ and $\boldsymbol{\eta}_2$ we have
$\frac{1}{\alpha}I(\boldsymbol{\eta}_1) < I(\boldsymbol{\eta}_2) < \alpha
I(\boldsymbol{\eta}_1)$, if $\frac{1}{\alpha} \boldsymbol{\eta}_1 <
\boldsymbol{\eta}_2 < \alpha \boldsymbol{\eta}_1$.

We first show that $I_k(\boldsymbol{\eta})$ is a standard interference
function, i.e., it  satisfies the following conditions:
\begin{center}
\begin{enumerate} [label=(\roman*)]
\item $I_k(\boldsymbol{\eta}) \geq 0, \forall \boldsymbol{\eta} \geq 0$,
\item $I_k(\boldsymbol{\eta}_1) \geq I_k(\boldsymbol{\eta}_2)$, if
$\boldsymbol{\eta}_1 \geq \boldsymbol{\eta}_2$,
\item For $\alpha >1, I_k(\alpha \boldsymbol{\eta}) < \alpha
I_k(\boldsymbol{\eta})$.
\end{enumerate}
\end{center}

It can be seen that $I_k(\boldsymbol{\eta}) \geq 0$ since both
$\boldsymbol{\eta}$ and ${\rm SINR}_k^u$ are positive.
To show that condition (ii) holds, we write
$I_k(\boldsymbol{\eta})$ in the form

\begin{equation}
\begin{aligned}
& I_k(\boldsymbol{\eta}) = \\
 & \frac{ \sum \limits_{m=1}^M \gamma_{mk} (
\frac{1}{\rho_u}+ \eta_k  \beta_{mk} ) +     \sum\limits_{i \neq k}
\eta_i
[ \sum\limits_{m=1}^M \beta_{mi}   \lVert {\bf a}_{m,k} \rVert^2_2 +  \tau
\rho_p ( | \sum\limits_{m=1}^M \beta_{mi}  \boldsymbol{\psi}^H_i {\bf
a}_{m,k} |^2+ \sum\limits_{m=1}^M \sum\limits_{j=1}^K \beta_{mi}
\beta_{mj} | \boldsymbol{\psi}^H_j {\bf a}_{m,k} |^2 )
]}{ ( \sum\limits_{m=1}^M \gamma_{mk} )^2}.
\end{aligned}
\label{EqAppB1}
\end{equation}

Now, let $\eta^{(1)}_k$ be $k$-th element of
$\boldsymbol{\eta}_1$.
Then we have
{
\small
\begin{equation}
\begin{aligned}
& I_k(\boldsymbol{\eta}_1) \\
& = \frac{ \sum \limits_{m=1}^M \gamma_{mk} ( \frac{1}{\rho_u}+
\eta^{(1)}_k  \beta_{mk} ) +     \sum\limits_{i \neq k} \eta_i^{(1)}
[ \sum\limits_{m=1}^M \beta_{mi}   \lVert {\bf a}_{m,k} \rVert^2_2 +  \tau
\rho_p ( | \sum\limits_{m=1}^M \beta_{mi}  \boldsymbol{\psi}^H_i {\bf
a}_{m,k} |^2+ \sum\limits_{m=1}^M \sum\limits_{j=1}^K \beta_{mi}
\beta_{mj} | \boldsymbol{\psi}^H_j {\bf a}_{m,k} |^2 )
]}{ ( \sum\limits_{m=1}^M \gamma_{mk} )^2}\\
&\geq \frac{ \sum \limits_{m=1}^M \gamma_{mk} ( \frac{1}{\rho_u}+
\eta^{(2)}_k  \beta_{mk} ) +     \sum\limits_{i \neq k} \eta_i^{(2)}
[ \sum\limits_{m=1}^M \beta_{mi}   \lVert {\bf a}_{m,k} \rVert^2_2 +  \tau
\rho_p ( | \sum\limits_{m=1}^M \beta_{mi}  \boldsymbol{\psi}^H_i {\bf
a}_{m,k} |^2+ \sum\limits_{m=1}^M \sum\limits_{j=1}^K \beta_{mi}
\beta_{mj} | \boldsymbol{\psi}^H_j {\bf a}_{m,k} |^2 )
]}{ ( \sum\limits_{m=1}^M \gamma_{mk} )^2}\\
&= I_k(\boldsymbol{\eta}_2),
\end{aligned}
\label{EqAppB2A}
\end{equation}
}
\normalsize
where the inequality holds due to linearity of the numerator term.

To show that condition (iii) holds, we notice that for $\alpha > 1$ we have

{ \small
\begin{equation}
\begin{aligned}
& I_k(\alpha \boldsymbol{\eta}) \\
&= \frac{ \sum \limits_{m=1}^M \gamma_{mk} ( \frac{1}{\rho_u}+  \alpha
\eta_k  \beta_{mk} ) +     \sum\limits_{i \neq k} \alpha \eta_i \left[
\sum\limits_{m=1}^M \beta_{mi}   \lVert {\bf a}_{m,k} \rVert^2_2 +  \tau \rho_p
\left( |\sum\limits_{m=1}^M \beta_{mi}  \boldsymbol{\psi}^H_i {\bf
a}_{m,k} |^2+ \sum\limits_{m=1}^M \sum\limits_{j=1}^K \beta_{mi}
\beta_{mj} | \boldsymbol{\psi}^H_j {\bf a}_{m,k} |^2 \right)
\right]}{ \left( \sum\limits_{m=1}^M \gamma_{mk} \right)^2}\\
& <  \frac{\alpha \sum \limits_{m=1}^M \gamma_{mk} ( \frac{1}{\rho_u}+
\eta_k  \beta_{mk} ) +     \sum\limits_{i \neq k} \alpha \eta_i \left[
\sum\limits_{m=1}^M \beta_{mi}   \lVert {\bf a}_{m,k} \rVert^2_2 +  \tau \rho_p
\left( |\sum\limits_{m=1}^M \beta_{mi}  \boldsymbol{\psi}^H_i {\bf
a}_{m,k} |^2+ \sum\limits_{m=1}^M \sum\limits_{j=1}^K \beta_{mi}
\beta_{mj} | \boldsymbol{\psi}^H_j {\bf a}_{m,k} |^2 \right)
\right]}{ \left( \sum\limits_{m=1}^M \gamma_{mk} \right)^2}\\
& = \alpha I_k(\boldsymbol{\eta}).
\end{aligned}
\label{EqAppB2}
\end{equation}
}
\normalsize
Now, let $\boldsymbol{\eta} = \frac{\boldsymbol{\eta}}{\alpha}$. Then, from
(iii), we get $\frac{1}{\alpha}I_k(\boldsymbol{\eta}) <
I_k(\frac{\boldsymbol{\eta}}{\alpha})$. If $\frac{1}{\alpha}
\boldsymbol{\eta}_1
< \boldsymbol{\eta}_2 < \alpha \boldsymbol{\eta}_1$, then

\[
\frac{1}{\alpha} I_k(\boldsymbol{\eta}_1) <
I_k(\frac{\boldsymbol{\eta}_1}{\alpha}) \stackrel{\text{(a)}}{<}
I_k(\boldsymbol{\eta}_2) \stackrel{\text{(b)}}{<} I_k(\alpha
\boldsymbol{\eta}_1) \stackrel{\text{(c)}}{<} \alpha I_k(\boldsymbol{\eta}_1),
\]
where (a) and (b) follow from condition (ii) and (c) follows from condition
(iii).
Thus,

\[
\frac{1}{\alpha} I_k(\boldsymbol{\eta}_1) < I_k(\boldsymbol{\eta}_2) < \alpha
I_k(\boldsymbol{\eta}_1),
\]

and $I_k(\boldsymbol{\alpha})$ is a two-sided scalable function. From this, we
can also show that

\[
\frac{1}{\alpha I_k(\boldsymbol{\eta}_1)}  < \frac{1}{I_k(\boldsymbol{\eta}_2)}
< \alpha \frac{1}{I_k(\boldsymbol{\eta}_1)}.
\]

Thus, $\frac{1}{I_k(\boldsymbol{\eta})}$ is also two-sided scalable.

\section*{Appendix C}
\label{AppC}
It is easy to see that
$$
\mathbb{E}[|T_1|^2] = \rho_d  \left( \sum_{m=1}^M
\eta_{mk}^{1/2} \gamma_{mk} \right)^2 \mbox{ and } \mathbb{E}[|{\rm T_4}|^2] = 1.
$$.

The variances of $T_2$ and $T_3$ are calculated similarly as in Appendix A.

\bibliographystyle{IEEEbib}
\bibliography{IEEEabrv,bibJournalList,refs}

\end{document}